\documentclass[twocolumn,showpacs,pre,floatfix]{revtex4}
\usepackage{graphics}
\usepackage{latexsym}
\usepackage{bm}

\begin{document}

\author{Peter Rupp, Reinhard Richter, and Ingo Rehberg}

\affiliation{Physikalisches Institut, Experimentalphysik V,
Universit\"at Bayreuth, D-95440 Bayreuth, Germany}

\date{\today}

\title{Critical exponents of directed percolation measured in spatiotemporal intermittency}

\begin{abstract}
A new experimental system showing a transition to spatiotemporal intermittency is presented. It consists of a ring of hundred oscillating ferrofluidic spikes. Four of five of the measured critical exponents of the system agree with those obtained from a theoretical model of directed percolation.
\end{abstract}


\maketitle

\section{Introduction}

Intermittency was observed in hydrodynamics as a precursor to turbulence (see e.g. \cite{tritton}). For dissipative dynamical systems without spatial degrees of freedom intermittency was first modeled by ordinary differential equations and iterative maps \cite{pomeau1980}. Spatiotemporal intermittency (STI) is a further development of this concept for spatially extended systems and was introduced by Kaneko \cite{kaneko1985} in the mid 80's \cite{test1}. It is characterized by patches of ordered and disordered states fluctuating stochastically in space and time.  This behaviour has been observed in many systems. Theoretical approaches have been made in a large variety of systems ranging from partial differential equations, such as the damped Kuramoto-Sivashinsky equation \cite{chate1987,frisch1986} or the complex Ginzburg-Landau equation \cite{daviaud1992,deissler1985,vanhecke1998}, over stochastical partial differential equations \cite{toral2000} to coupled map lattices (CML) \cite{kaneko1985,chate1988a} and probabilistic cellular automata \cite{chate1989}. 

In 1986 Pomeau \cite{pomeau1986} suggested that the onset of chaos via STI might be analogous to directed percolation (DP) \cite{test2}. Such processes are modeled as a probabilistic cellular automaton with two states per site, associated with the laminar and chaotic patches in the case of STI. One of the main features of DP is the presence of an absorbing state, which corresponds to the laminar state. The absorbing state prevents the nucleation of chaotic domains within laminar domains. DP model predicts some universal properties of STI. In particular, the fraction of chaotic domains is expected to grow with a power law $\epsilon^{\beta}$, where $\epsilon$ measures the distance from threshold. The correlation length decreases with $\epsilon^{-\nu_{\rm{s}}}$, as does the correlation time with $\epsilon^{-\nu_{\rm{t}}}$, and the critical distribution of the laminar lengths is determined by $l^{-\mu_{\rm{s}}}$ and by $t^{-\mu_{\rm{t}}}$ for the laminar times.
\begin{table*}
\caption{Experiments and results in quasi onedimensional.}
\begin{tabular}{lclcclcccc}
Authors          & Year & Experiment & Size & $T_0 (s)$ & Geometry & $\beta$ & $\nu_s$ & $\nu_t$ & $\mu_s$ \\
\hline
Ciliberto et al. \cite{ciliberto1992} & 1988 & RB-convection       & 20      & 10   & annular & --           & $0.5$         & -- & $1.9\pm0.1$ \\
Daviaud et al. \cite{daviaud1990,daviaud1992}& 1990 & RB-convection & 40     & 2    & linear  & $0.3\pm0.05$ & $0.5\pm0.05$  & $0.5\pm0.05$ & $1.6\pm0.2$ \\
~~~~~~~~"                             & "    & ~~~~~~~~~~"         & 30      & 2    & annular & --           & $0.5$         & $0.5$ & $1.7\pm0.1$ \\
Michalland et al.\cite{michalland1993}& 1993 & viscous fingering   & 40      & 1.5  & linear  & $0.45\pm0.05$& $0.5$         & -- & $0.63\pm0.02$ \\
Willaime et al. \cite{willaime1993}   & 1993 & line of vortices    & 15      & 5    & linear  & --           & --            & $0.5$ & -- \\
Degen et al. \cite{degen1996}         & 1996 & Taylor-Dean         & 20 (90) & 1.5  & linear  & $1.30\pm0.26$& $\approx0.64$ & $\approx0.73$ & $1.67\pm0.14$\\
Colovas et al. \cite{colovas1997}     & 1997 & Taylor-Couette      & 30 (70) & 0.5  & linear  & --           & --            & -- & --\\
Bottin et al. \cite{bottin1997}       & 1997 & plane Couette       &  --     & --   & linear  & --           & --            & -- & --\\
Vallette et al. \cite{gollub1997}     & 1997 & fluid fronts        & 40      & 0.5  & linear  & --           & --            & -- & --\\  
Jensen (theory) \cite{jensen1999,hinrichsen2000a}       & 1999 & directed percolation & --     & --   & --      & $0.276486(8)$      & $1.096854(4)$       & $1.733847(6)$      & $1.748$\\
present paper                         & --   & ferrofluidic spikes & 108     & 0.08 & annular & $0.3\pm0.05$ & $1.2\pm0.1$   & $0.7\pm0.05$ & $1.7\pm0.05$ 
\end{tabular}
\label{table01}
\end{table*}

Some experimental tests of the conjecture \cite{pomeau1986} in quasi one-dimensional systems have been made \cite{ciliberto1992,daviaud1990,daviaud1992,michalland1993,willaime1993,degen1996,colovas1997,bottin1997,gollub1997}. A short summary of these and the relevant exponents \cite{jensen1999,hinrichsen2000a} for the comparison with the DP-model are given in Table\,\ref{table01}. However the statement that "...there is still no experiment where the critical behaviour of DP was seen" by Grassberger \cite{grassberger1997} still seems to be true. Thus in this paper a new experimental approach to this old problem is presented. 

The system introduced consists of a ring of ferrofluidic spikes excited by an external magnetic field. Ferrofluids, also called magnetic fluids, are a collodial suspension of ferromagnetic nanoparticels. The fluid is superparamagnetic \cite{rosensweig}. The idea for using this fluid was motivated by the fact that a single peak of ferrofluid can show chaotic oscillations under external driving of a magnetic field \cite{mahr1998,friedrichs2000}. We introduce here a system where about $100$ of these oscillating peaks are coupled by magnetic and hydrodynamic interaction. They exhibit changes in peak height of about $10\,\%$ and variations in wavelength $\lambda$ of about $50\,\%$. This system is advantageous because of its short response times and the easy control of the excitation.

The article is organized as follows: In Sec.\,\ref{expsetup} we describe the experimental setup, the procedure of the measurement and the data extraction methods. In Sec.\,\ref{res} the quantitative results are presented, i.e. the critical exponents $\beta$, $\nu_s$, $\nu_t$, $\mu_s$ and $\mu_t$. Finally in Sec.\,\ref{disc} the results are discussed and an outlook to further investigations is given.

\section{\label{expsetup}Experiment}

\begin{figure}[h]
\includegraphics{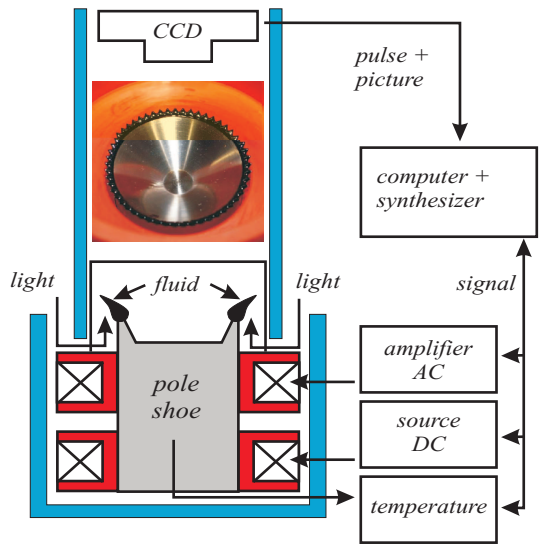}
\caption{\label{setup}Sketch of the experimental setup.}
\end{figure}

The experiment is based on the Rosensweig instability \cite{rosensweig}. This instability is observed in a horizontal layer of magnetic fluid, when a threshold of the vertically oriented magnetic field is surpassed. The flat surface becomes unstable and a pattern of liquid spikes emerges. In case of an inhomogenous field the wavelength of this pattern can be controlled by the magnetic field $H$. The wavelength scales with the gradient of the magnetic field divided by a basic field at the undisturbed surface of the fluid. A larger gradient emphasizes small peaks and suppresses large ones. Thus, inorder to generate as many spikes as possible, the setup of the experiment consists of a cylindrical electromagnet with a sharp edge (Fig.\,\ref{setup}). The magnetic fluid is trapped by the inhomogenous magnetic field at this edge of the magnetically soft iron core. In that way the $40\,\textrm{mm}$ diameter of the pole shoe supports a ring of up to $130$ spikes of magnetic fluid as indicated by the picture in Fig.\,\ref{setup}.

The magnetic fluid used in this experiment is EMG901 from $Ferrofluidics^{TM}$, a fluid based on magnetite $Fe_3O_4$ with isoparafin as carrier fluid. At $20^{\circ}\,\textrm{C}$ EMG901 has a density $\rho=1.53\,\textrm{kg}\cdot\textrm{m}^{-3}$, a surface tension $\sigma=29\,\textrm{mN}\cdot\textrm{m}^{-1}$, a dynamical viscosity $\eta=25\,\textrm{mPas}$ and a susceptibility $\xi=3$. 

The ring of spikes is recorded with a CCD-camera mounted above the pole shoe. The electromagnet consists of a bias coil and an excitation coil. The bias coil is provided with a direct current of $I=1.0\,\textrm{A}$ to keep the magnetic fluid in its place. The excitation coil is driven by an alternating current, phase-locked with the camera frequency, providing a stroboscopic jitter free recording on long timescales. The alternating current can be adjusted between $0$ and $4.1\,\textrm{A}$. In this interval the number of spikes ranges from $60$ to $130$. For the amplitudes used in the experiment $108$ spikes have been observed.

To keep the viscosity and the surface tension constant the fluid is temperature controlled to $12.5\pm0.03\,^{\circ}\textrm{C}$ by cooling the pole shoe. To prevent the evaporation of the isoparafin, the volume around the egde of the pole shoe is sealed with a glass plate to provide long term stability.

\begin{figure}[h]
\includegraphics{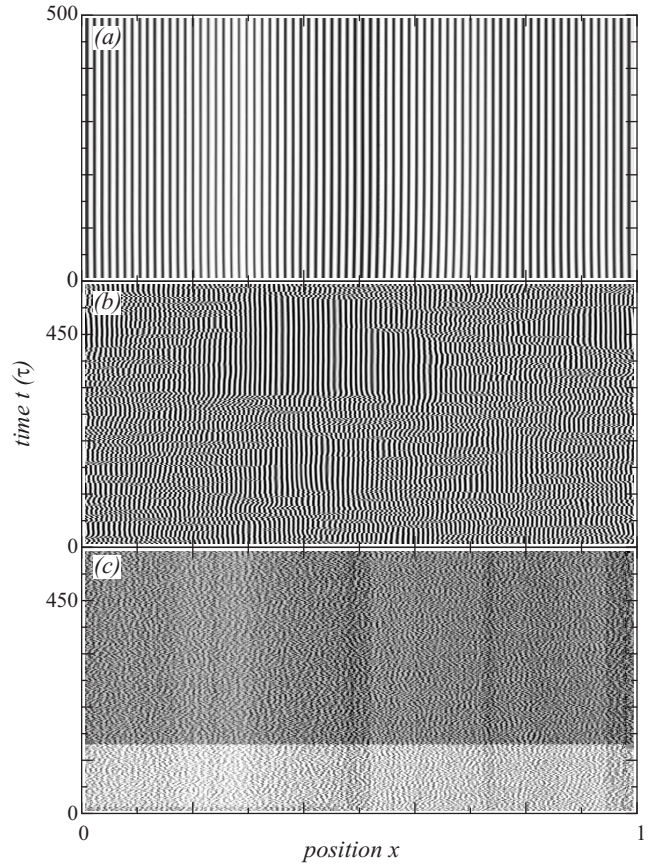}
\caption{\label{states}Space time plots of the different states of the system. (a) $I_{\rm{ex}}=2.8\,\textrm{A}$: laminar state; (b) $I_{\rm{ex}}=3.0\,\textrm{A}$: spatiotemporal intermittency; (c) $I_{\rm{ex}}=3.6\,\textrm{A}$: chaotic state. $500$ excitation periods are shown. The position $x$ is normalized over the size of the ring.}
\end{figure}
\begin{figure*}[ht]
\includegraphics{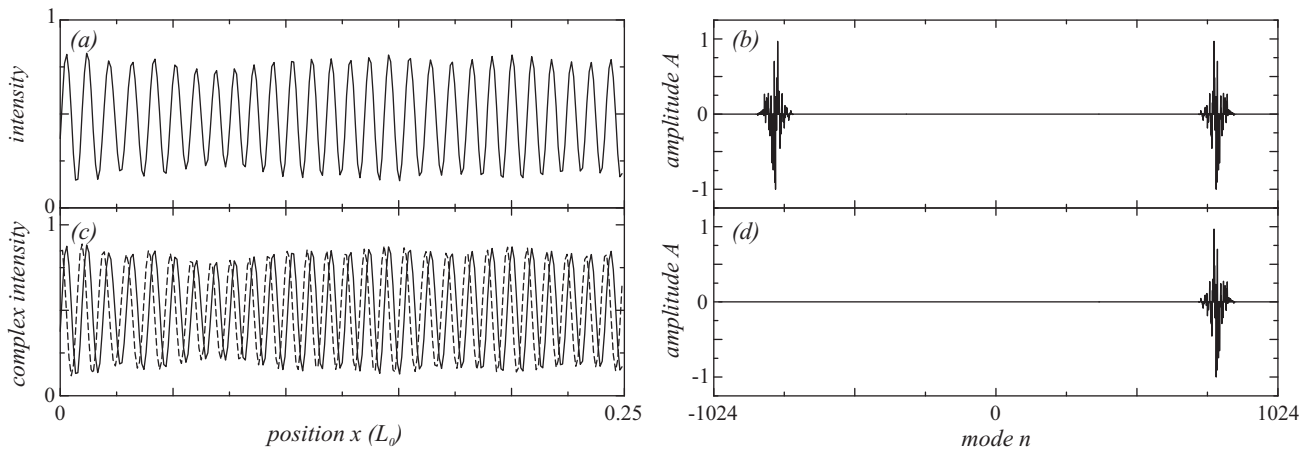}
\caption{\label{complex}Complexification of the data (see text). The left hand side displays the data in real space ((a),(c)), the rigth hand side the corresponding graphs in Fourier space ((b),(d)).}
\end{figure*}

The spatiotemporal behaviour is investigated by observing the ring with the CCD-camera. To extract the wavelength and amplitude of the spikes as a function of space, we define a ring of interest around the center of the pole shoe which is covering the ring of spikes. It is divided into $1024$ segments. The average of the greyvalues within each segment represents the amplitude. That way we get a spatial resolution of around $10$ segments  per peak. This reduction of the $2D$ image to a single line scan can be done in real time with a frequency of $12.5\,\textrm{Hz}$.

In Fig.\,\ref{states}\,(a) $500$ of such scans of a laminar state are shown in space and time, where dark regions correspond to high amplitudes. The driving frequency of the excitation field is $f_{\rm{ex}}=12.5\,\textrm{Hz}$ as mentioned above. The period $\tau=1/f_{\rm{ex}}$ is used to scale the time. Due to this stroboscopic recording the oscillations of the spikes cannot be seen. 

For the measurements a current of $1\,\textrm{A}$ is applied to the bias coil and a constant volume of the magnetic fluid is dropped on the egde of the pole shoe. After a waiting time of $2\,\textrm{h}$ thermal equilibrium is reached. Then the second coil is provided with a sinusodial excitation signal of an amplitude of $I_{\rm{ex}} = 4.05\,\textrm{A}$, driving the system into the fully chaotic regime similar to the one indicated in Fig.\,\ref{states}\,(c). That state is the basis for a quench to a lower excitation value corresponding to the STI regime (Fig.\,\ref{states}\,(b)), which is then analyzed subsequently. The recording of the data starts after a waiting time of $\approx 1800\tau$, when the transients following the quench have died out. The data are recorded for $2000$ excitation periods $\tau$. For higher excitation amplitudes the laminar state (Fig.\,\ref{states}\,(a)) becomes intermittent in space and time (Fig.\,\ref{states}\,(b)) and eventually chaotic (Fig.\,\ref{states}\,(c)).   

\begin{figure}[h]
\includegraphics{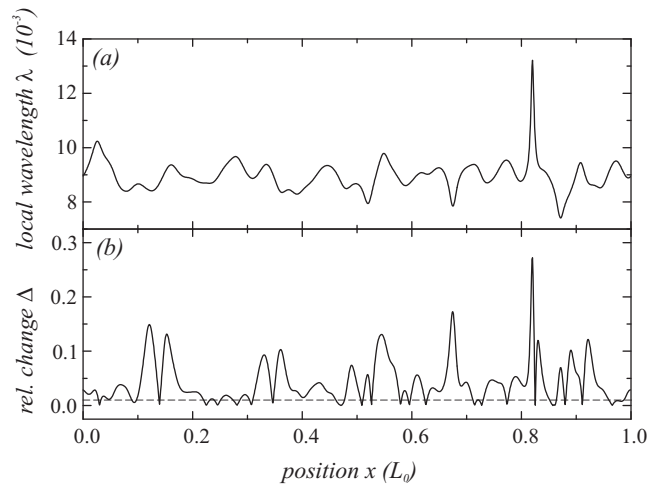}
\caption{\label{lambda}(a) Local wavelength $\lambda$ calculated from Fig,\,\ref{complex}\,(c). (b) Relative change of local wavelength $\Delta$. The dashed line corresponds to the threshold of $0.01$ taken from Fig.\,\ref{dldistrib}.}
\end{figure}

\begin{figure}[h]
\includegraphics{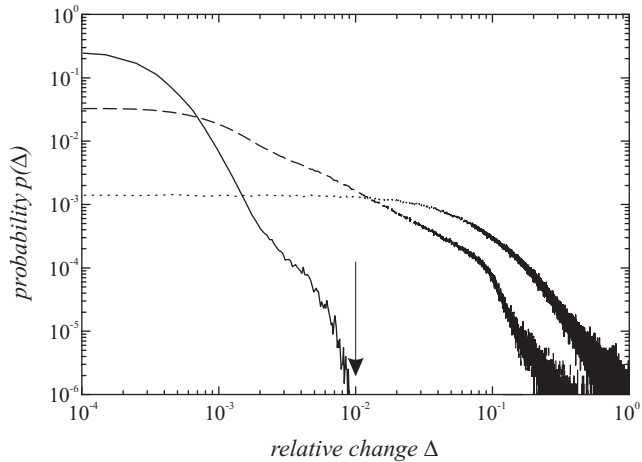}
\caption{\label{dldistrib}Distributions of the relative change of wavelength $\Delta$ for three different $I_{\rm{ex}}$: solid line $I_{\rm{ex}}=2.8\,\textrm(A)$ corresponding to Fig.\,\ref{states}\,(a), dashed line $I_{\rm{ex}}=3.0\,\textrm(A)$ corresponding to Fig.\,\ref{states}\,(b) and dotted line $I_{\rm{ex}}=3.5\,\textrm(A)$ corresponding to Fig.\,\ref{states}\,(c). The cutoff of the distribution for the fully laminar state gives the threshold of the binarization $\Delta_{\rm{c}}=0.01$, as depicted by the arrow.}
\end{figure}

\begin{figure}
\includegraphics{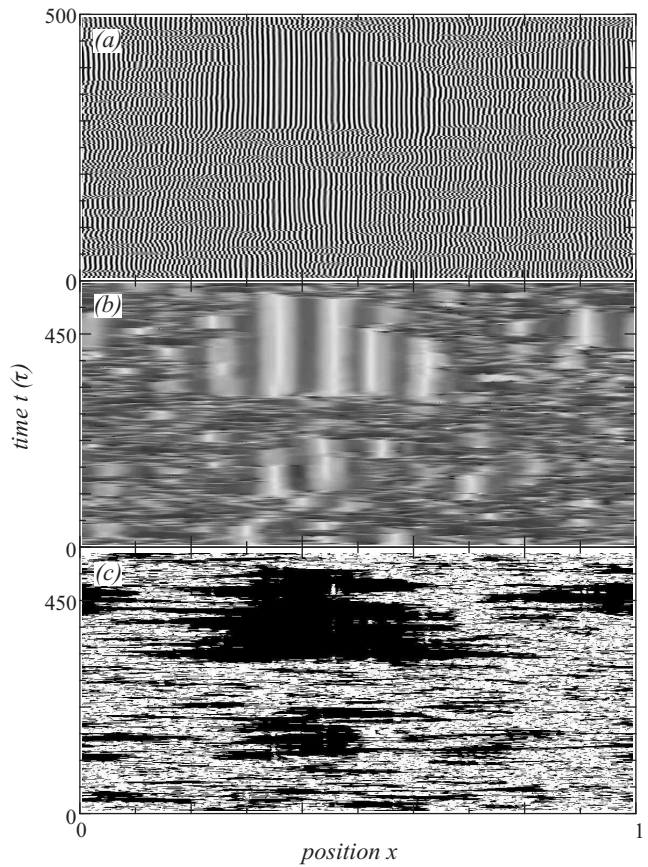}
\caption{\label{steps}x-t-plots at different states of the data processing. (a) raw data; (b) same section as in (a) after transforming the data. White corresponds to large local wavelengths, black to small ones. (c) final step: the relative change of local wavelength over two subsequent periods $\Delta$ is calculated and binarized. The black areas are defined as regular ($\Delta<0.01$), the white are chaotic ($\Delta>0.01$).}
\end{figure}

For a quantitative analysis of the transition a reliable detection of the regular domains is fundamental. Different criteria for the distinction of the regular and chaotic domains have been proposed \cite{ciliberto1992} -\cite{gollub1997}. Because of the strong variations of the wavelength of our system we have selected a wavelength criterion, which is based on the local wavelenght $\lambda(x,t)$. To obtain $\lambda(x,t)$ we use a method called \textit{complexification}. By this method a zero imaginary part is added to every real value (see Fig.\,\ref{complex}\,(a)). Then the data are transformed to Fourier space, where the amplitudes corresponding to negative wavenumbers are eliminated (Fig.\,\ref{complex}\,(b),(d)). With a backward transformation to real space, every value has a non zero imaginary part (Fig.\,\ref{complex}\,(c) dashed line). $\lambda$ is subsequently calculated as the phase difference between neighbouring values in real space (Fig.\,\ref{lambda}\,(a)). In the last step the relative change of the local wavelength 
\begin{equation}
\Delta=\frac{\vert\lambda_{\rm{t+1}}-\lambda_{\rm{t}}\vert}{\lambda_{\rm{t}}}
\end{equation}
is calculated. To get a clear distinction between regular and irregular domains, changes in $\Delta$ which are larger than $0.01$ are counted as irregular, whereas smaller changes belong to regular domains (Fig.\,\ref{lambda}\,(b)). The threshold value $\Delta_{\rm{c}}$ is derived from the distribution of $\Delta$ for the fully laminar case  which is presented in Fig.\,\ref{dldistrib} by a solid line. There are no larger variations than $1\%$. Smaller values of $\Delta$ are artefacts of the recording technique and thus are suppressed. For comparison the dashed (dotted) lines give the distribution of $\Delta$ for the intermittent (chaotic) states respectively. The calculations of the exponents are robust to changes of the threshold of up to $50\%$. This variation of the binarization threshold corresponds to changes in the resulting critical exponents within the statistical errors. In Fig.\,\ref{steps} we demonstrate the application of the above described procedure to the intermittent data. Fig.\,\ref{steps}\,(a) gives the raw data, (b) displayes the local wavelength and (c) the relative change $\Delta$ after binarization. The quantitative analysis of STI described below is based on these binarized information. To get a better signal-noise-ratio we average the results over six independent runs of the experiment, which include six refills of the apparatus with fresh fluid.

\section{\label{res}Experimental Results}

At low excitation amplitudes $I_{\rm{ex}} \ll 3.0\,\textrm{A}$ the system is completely regular (Fig.\,\ref{states} (a)) showing $108$ spikes. Slight spatial variations of the wavelength $\Delta < 0.01$ remain constant in time. In Fig.\,\ref{states} (b) at $I_{\rm{ex}}=3.0\,\textrm{A}$ irregular fluctuations are apparent, which we consider as a manifestation of STI. A fairly clear distinction between regular and irregular domains can be made in this image even by naked eye. Further increase of $I_{\rm{ex}}$ leads to a spreading of the irregular domains engulfing the regular regions, until finally the whole system is chaotic (Fig.\,\ref{states} (c)).

\begin{figure}
\includegraphics{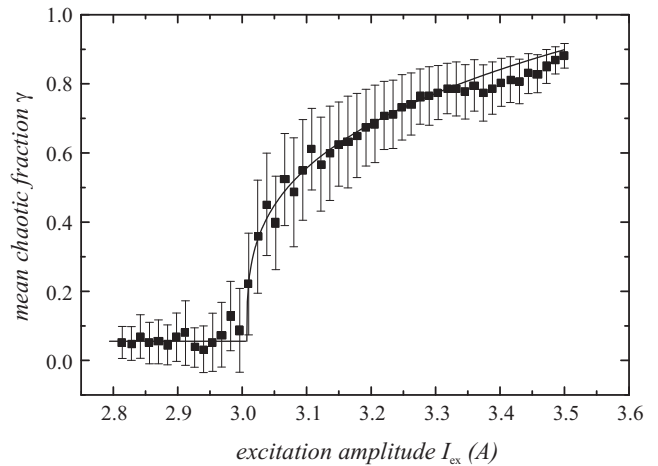}
\caption{\label{cf}The mean chaotic fraction $\gamma$ vs. excitation amplitude $I_{\rm{ex}}$. The solid line is a power law fit. The error bars represent the statistical errors.}
\end{figure}

As an order parameter for STI we take the \textit{time-averaged chaotic fraction} $\gamma$, which is the ratio of chaotic regions to the length of the system. This ratio is averaged over the $2000$ excitation periods $\tau$. Its variation with the control parameter $I_{\rm{ex}}$ is shown in Fig.\,\ref{cf}. The error bars represent the variance of the chaotic fraction.

Close to the onset of STI the mean chaotic fraction is expected to grow with a power law 
\begin{equation}
\gamma\sim(I_{\rm{ex}}-I_{\rm{c}})^{\beta}.
\end{equation}
The solid line in Fig.\,\ref{cf} is a fit to our data, using $I_{\rm{c}}$, $\beta$, and an offset representing background noise as adjustable parameters. The threshold value determined in this way is $I_{\rm{c}}=3.0\pm0.05\,\textrm{A}$ and the exponent $\beta = 0.3\pm0.05$ is in agreement with the theoretical expectation for DP $\beta=0.276486(8)$ \cite{hinrichsen2000a}. 
\begin{figure}
\includegraphics{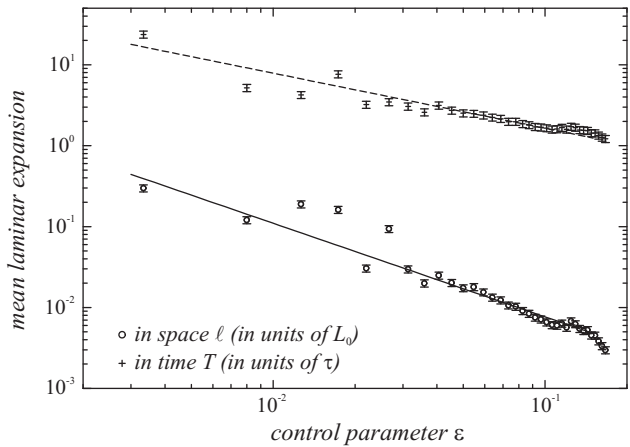}
\caption{\label{mll}The mean laminar expansion in space $\ell$ and in time $T$ vs. control parameter $\epsilon$. The lines are power law fits. The error bars represent statistical errors.}
\end{figure}

Another way to characterize the regular domains is to look at the mean laminar expansion in space $\ell$ and time $T$. First we define the laminar lengths $l$ as the number of consecutive regular segments between two chaotic ones devided by the total number of segments at a certain time $t_0$. The laminar times $t$ are the number of segments between two chaotic ones at a certain position $x_0$. The averages of these numbers are displayed in Fig.\,\ref{mll} as the function of the normalized controlparameter 
\begin{equation}
\epsilon = \frac{I_{\rm{ex}}}{I_{\rm{c}}}-1
\end{equation}
Both parameters decay with a power law:
\begin{eqnarray}
\ell \sim \epsilon^{-\nu_{\rm{s}}^*}\label{nus}\\
T \sim \epsilon^{-\nu_{\rm{t}}^*}\label{nut}
\label{eqnu}
\end{eqnarray}
and the fit yields the exponents $\nu_{\rm{s}}^* = 1.2\pm0.2$,which is represented by the solid line and $\nu_{\rm{t}}^* = 0.7\pm0.1$, which is represented by the dashed line. Only data in the range of $0.03<\epsilon<0.1$ are taken into account. For smaller $\epsilon$ a finite size effect is obvious: Following Cross and Hohenberg \cite{cross1993} the characteristic length of the regular domains  $\ell$ must be much smaller than the system size $1$. For $\epsilon>0.1$ the system is no longer intermittent, but rather chaotic. 
\begin{figure}
\includegraphics{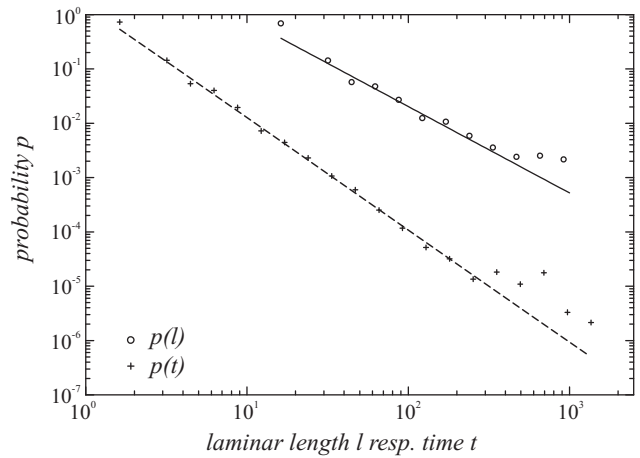}
\caption{\label{distribution01}Distribution of the laminar domain length $l$ and time $t$ for $\epsilon=0$. The solid lines are power law fits. To suppress the statistical fluctuations the distributions are logarithmical binned.}
\end{figure}

In Fig.\,\ref{distribution01} the distributions of the laminar domain length and time for $\epsilon = 0$ are presented. At the threshold the distribution should follow a power law for both the distribution of the laminar domain lengths $l$ and times $t$. To suppress the statistical fluctuations the values are logarithmically binned. The solid line represent the power law fit for the distribution of the lengths (\ref{mus}) and for the times (\ref{mut}):
\begin{eqnarray}
p(l) \sim l^{-\mu_{\rm{s}}}\label{mus},\\
p(t) \sim l^{-\mu_{\rm{t}}}\label{mut}.
\end{eqnarray}
We obtain $\mu_{\rm{s}}=1.7\pm0.05$, in agreement with the theoretical value $\mu_{\rm{s}}=1.734$ and $\mu_{\rm{t}}=2.1\pm0.1$ in accordance with the theoretical value $\mu_{\rm{t}}>2.0$. 

\begin{figure}
\includegraphics{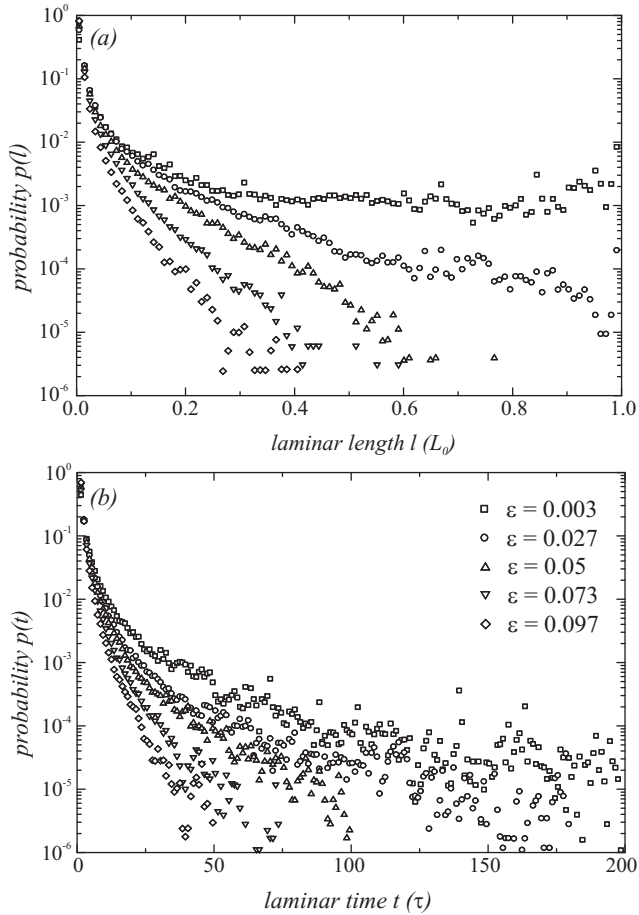}
\caption{\label{distribution02}Distributions of the laminar length $l$ (a) and the laminar time $t$ (b) for $\epsilon>0$. The legend in (b) holds for both figures.}
\end{figure}

For $\epsilon>0$ the power law has a cutoff at the correlation length resp. time of the system and an exponential tail with a decay length equal to the correlation length $\xi_{\rm{decay}}$ resp. time  $\theta_{\rm{decay}}$ (Fig.\,\ref{distribution02}).  

\begin{figure}
\includegraphics{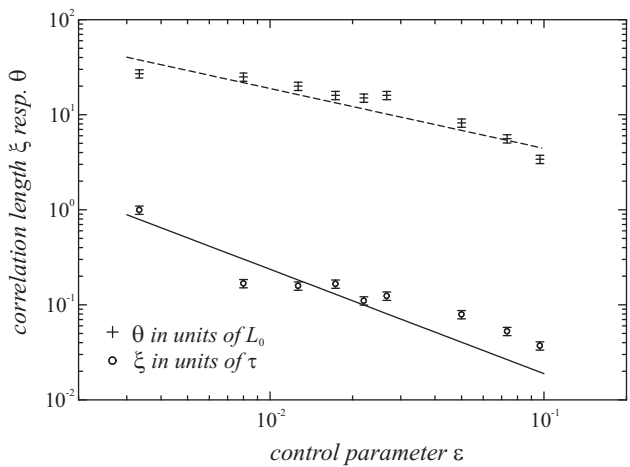}
\caption{\label{decay}The correlation length $\xi$ and time $\theta$ derived from the exponential decay of the distributions (Fig.\,\ref{distribution02}) vs. control parameter $\epsilon$. The lines are power law fits. The error bars represent statistical errors.}
\end{figure}

Both parameters should grow with the same power law as the mean expansions $\ell$ and $T$ in Eqs.\,(\ref{nus}) and (\ref{nut}). This behaviour can be seen in Fig.\,\ref{decay}: The lines correspond to power law fits with the exponents $\nu_{\rm{s}} = 1.1\pm0.2$ and $\nu_{\rm{t}} = 0.62\pm0.14$. The large errors are due to the statistical fluctuations of the distributions and difficulties in the definition of the cutoff length.
\begin{figure}
\includegraphics{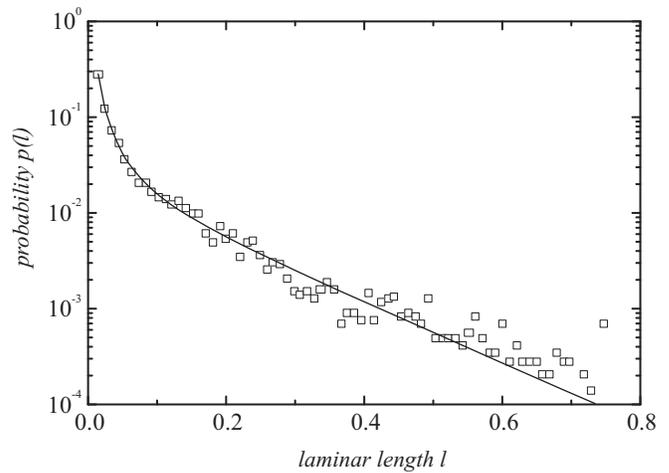}
\caption{\label{distribution03}Distribution of the laminar domain length $l$ for $\epsilon=0.019$. The solid line stems from Eq.\,(\ref{eq_powexp}).}
\end{figure}
 
The distributions cannot be described by a simple power law. A more complicated distribution function has been suggested 
\begin{equation}
p(l)=(Al^{-\mu}+B)e^{-l/l_{\rm{decay}}}
\label{eq_powexp}
\end{equation}
in Ref.\,\cite{ciliberto1992}. The solid line in Fig.\,\ref{distribution03} is a fit to this empirical distribution function for $\epsilon=0.019$ with $l_{\rm{decay}}=0.17$. It shows clearly that the power law is now replaced by a function more reminiscent of an exponential decay.

\section{\label{disc}Discussion}

To conclude, we have presented a new experimental system exhibiting STI. In contrast to all previous experiments, which are autonomous ones, our system is periodically driven. We have measured the critical exponents $\beta$, $\nu_{\rm{s}}$, $\nu_{\rm{t}}$, $\mu_{\rm{s}}$ and $\mu_{\rm{t}}$ for the mean chaotic fraction $\gamma$, the mean laminar length $\ell$, the mean laminar time $T$, the correlation length $\xi_{\rm{decay}}$ and time $\theta_{\rm{decay}}$ and the laminar length and time distribution functions for $\epsilon\approx0$. Four of the five exponents agree with the theoretical expectation derived from a DP model within our experimental resolution. Considering the simplicity of the underlying discrete model, the fact that the theory is applicable only near $I_{\rm{c}}$, the complexity of our experiment and the fact that our apparatus has a finite size this concordance seems truly remarkable.

The fact that the parameter $\nu_{\rm{t}}$ is far from the expected value needs further discussion. One difference between the underlying model and our experiment might be the nature of the absorbing state. In fact, the absence of a truly absorbing state seems to be common to all experimental results so far \cite{kaneko}. Following the ideas presented in that book the allowance for a continuum of states between the absorbing and the active one might result in a more realistic description of the experimental situation.

Another way to soften the assumption of a truly absorbing state is the introduction of stochastic mechanisms permitting the nucleation of chaotic domains. In DP this is done by implementing a weak external field which creates chaotic domains in a pure laminar neighbourhood with a certain probability \cite{janssen1999,hinrichsen2000a}. As long as the probability of the creation processes is small the critical behaviour of DP is only slightly disturbed. If the probability becomes too large the universality is destroyed. The stochastic field could be an anologon to the "background noise" of our experimental apparatus.

Alternatively, traveling soliton-like structures have added to CML and DP \cite{bohr2001}. The interaction of these structures might lead to chaotic domains nucleating in laminar regions, which again softens the assumption of a truly absorbing state and leads to the breakdown of universality in DP.

If any of these ideas will be able to explain the behaviour of the measured decay times is subject to further investigation.

\begin{acknowledgments}
The authors would like to thank Hugues Chat\'e and Haye Hinrichsen for helpful discussions. One of us (R. R.) would like to thank Victor Steinberg for inspiring contributions. The experiments have been financially supported by Deutsche Forschungsgemeinschaft through Re588/12.
\end{acknowledgments}


\begin{thebibliography}{32}
\expandafter\ifx\csname natexlab\endcsname\relax\def\natexlab#1{#1}\fi
\expandafter\ifx\csname bibnamefont\endcsname\relax
  \def\bibnamefont#1{#1}\fi
\expandafter\ifx\csname bibfnamefont\endcsname\relax
  \def\bibfnamefont#1{#1}\fi
\expandafter\ifx\csname citenamefont\endcsname\relax
  \def\citenamefont#1{#1}\fi
\expandafter\ifx\csname url\endcsname\relax
  \def\url#1{\texttt{#1}}\fi
\expandafter\ifx\csname urlprefix\endcsname\relax\def\urlprefix{URL }\fi
\providecommand{\bibinfo}[2]{#2}
\providecommand{\eprint}[2][]{\url{#2}}

\bibitem[{\citenamefont{Tritton}(1988)}]{tritton}
\bibinfo{author}{\bibfnamefont{D.~J.} \bibnamefont{Tritton}},
  \emph{\bibinfo{title}{Physical Fluid Dynamics}}
  (\bibinfo{publisher}{Clarendon Press and Oxford}, \bibinfo{year}{1988}),
  \bibinfo{edition}{2nd} ed.

\bibitem[{\citenamefont{Pomeau and Manneville}(1980)}]{pomeau1980}
\bibinfo{author}{\bibfnamefont{Y.}~\bibnamefont{Pomeau}} \bibnamefont{and}
  \bibinfo{author}{\bibfnamefont{P.}~\bibnamefont{Manneville}},
  \bibinfo{journal}{Commun.\ Math.\ Phys.} \textbf{\bibinfo{volume}{74}},
  \bibinfo{pages}{189} (\bibinfo{year}{1980}).

\bibitem[{\citenamefont{Kaneko}(2000)}]{kaneko1985}
\bibinfo{author}{\bibfnamefont{K.}~\bibnamefont{Kaneko}},
  \bibinfo{journal}{Prog.\ Theor.\ Phys.} \textbf{\bibinfo{volume}{74}},
  \bibinfo{pages}{1033} (\bibinfo{year}{2000}).

\bibitem{test1}
For a more general review see \protect{\cite{manneville}} and \protect{\cite{kaneko}}.

\bibitem[{\citenamefont{Chat\'e and Manneville}(1987)}]{chate1987}
\bibinfo{author}{\bibfnamefont{H.}~\bibnamefont{Chat\'e}} \bibnamefont{and}
  \bibinfo{author}{\bibfnamefont{P.}~\bibnamefont{Manneville}},
  \bibinfo{journal}{Phys.\ Rev.\ Lett.} \textbf{\bibinfo{volume}{58}},
  \bibinfo{pages}{112} (\bibinfo{year}{1987}).

\bibitem[{\citenamefont{Frisch et~al.}(1986)\citenamefont{Frisch, She, and
  Thual}}]{frisch1986}
\bibinfo{author}{\bibfnamefont{U.}~\bibnamefont{Frisch}},
  \bibinfo{author}{\bibfnamefont{Z.~S.} \bibnamefont{She}}, \bibnamefont{and}
  \bibinfo{author}{\bibfnamefont{O.}~\bibnamefont{Thual}},
  \bibinfo{journal}{J.\ Fluid\ Mech.} \textbf{\bibinfo{volume}{168}},
  \bibinfo{pages}{221} (\bibinfo{year}{1986}).

\bibitem[{\citenamefont{Daviaud et~al.}(1992)\citenamefont{Daviaud, Lega,
  Berge, Coullet, and Dubois}}]{daviaud1992}
\bibinfo{author}{\bibfnamefont{F.}~\bibnamefont{Daviaud}},
  \bibinfo{author}{\bibfnamefont{J.}~\bibnamefont{Lega}},
  \bibinfo{author}{\bibfnamefont{P.}~\bibnamefont{Berge}},
  \bibinfo{author}{\bibfnamefont{P.}~\bibnamefont{Coullet}}, \bibnamefont{and}
  \bibinfo{author}{\bibfnamefont{M.}~\bibnamefont{Dubois}},
  \bibinfo{journal}{Phys.\ Rev.\ Lett.} \textbf{\bibinfo{volume}{55}},
  \bibinfo{pages}{287} (\bibinfo{year}{1992}).

\bibitem[{\citenamefont{Deissler}(1985)}]{deissler1985}
\bibinfo{author}{\bibfnamefont{R.~J.} \bibnamefont{Deissler}},
  \bibinfo{journal}{J.\ Stat.\ Phys.} \textbf{\bibinfo{volume}{40}},
  \bibinfo{pages}{371} (\bibinfo{year}{1985}).

\bibitem[{\citenamefont{van Hecke}(1998)}]{vanhecke1998}
\bibinfo{author}{\bibfnamefont{M.}~\bibnamefont{van Hecke}},
  \bibinfo{journal}{Phys.\ Rev.\ Lett.} \textbf{\bibinfo{volume}{80}},
  \bibinfo{pages}{1896} (\bibinfo{year}{1998}).

\bibitem[{\citenamefont{Zimmermann et~al.}(2000)\citenamefont{Zimmermann,
  Toral, Piro, and Miguel}}]{toral2000}
\bibinfo{author}{\bibfnamefont{M.~G.} \bibnamefont{Zimmermann}},
  \bibinfo{author}{\bibfnamefont{R.}~\bibnamefont{Toral}},
  \bibinfo{author}{\bibfnamefont{O.}~\bibnamefont{Piro}}, \bibnamefont{and}
  \bibinfo{author}{\bibfnamefont{M.~S.} \bibnamefont{Miguel}},
  \bibinfo{journal}{Phys.\ Rev.\ Lett.} \textbf{\bibinfo{volume}{85}},
  \bibinfo{pages}{3612} (\bibinfo{year}{2000}).

\bibitem[{\citenamefont{Chat\'e and Manneville}(1988)}]{chate1988a}
\bibinfo{author}{\bibfnamefont{H.}~\bibnamefont{Chat\'e}} \bibnamefont{and}
  \bibinfo{author}{\bibfnamefont{P.}~\bibnamefont{Manneville}},
  \bibinfo{journal}{Physica D} \textbf{\bibinfo{volume}{32}},
  \bibinfo{pages}{409} (\bibinfo{year}{1988}).

\bibitem[{\citenamefont{Chat\'e and Manneville}(1989)}]{chate1989}
\bibinfo{author}{\bibfnamefont{H.}~\bibnamefont{Chat\'e}} \bibnamefont{and}
  \bibinfo{author}{\bibfnamefont{P.}~\bibnamefont{Manneville}},
  \bibinfo{journal}{Physica D} \textbf{\bibinfo{volume}{37}},
  \bibinfo{pages}{33} (\bibinfo{year}{1989}).

\bibitem[{\citenamefont{Pomeau}(1986)}]{pomeau1986}
\bibinfo{author}{\bibfnamefont{Y.}~\bibnamefont{Pomeau}},
  \bibinfo{journal}{Physica D} \textbf{\bibinfo{volume}{23}},
  \bibinfo{pages}{3} (\bibinfo{year}{1986}).

\bibitem{test2}
A good introduction to DP is found in Ref.\,\cite{kinzel1983}, for a more recent review of most aspects of DP see Ref.\,\cite{hinrichsen2000a}.

\bibitem[{\citenamefont{Caponeri and Ciliberto}(1992)}]{ciliberto1992}
\bibinfo{author}{\bibfnamefont{M.}~\bibnamefont{Caponeri}} \bibnamefont{and}
  \bibinfo{author}{\bibfnamefont{S.}~\bibnamefont{Ciliberto}},
  \bibinfo{journal}{Physica\ D} \textbf{\bibinfo{volume}{58}},
  \bibinfo{pages}{365} (\bibinfo{year}{1992}).

\bibitem[{\citenamefont{Daviaud et~al.}(1990)\citenamefont{Daviaud, Bonetti,
  and Dubois}}]{daviaud1990}
\bibinfo{author}{\bibfnamefont{F.}~\bibnamefont{Daviaud}},
  \bibinfo{author}{\bibfnamefont{M.}~\bibnamefont{Bonetti}}, \bibnamefont{and}
  \bibinfo{author}{\bibfnamefont{M.}~\bibnamefont{Dubois}},
  \bibinfo{journal}{Phys.\ Rev.\ A} \textbf{\bibinfo{volume}{42}},
  \bibinfo{pages}{3388} (\bibinfo{year}{1990}).

\bibitem[{\citenamefont{Michalland et~al.}(1993)\citenamefont{Michalland,
  Rabaud, and Couder}}]{michalland1993}
\bibinfo{author}{\bibfnamefont{S.}~\bibnamefont{Michalland}},
  \bibinfo{author}{\bibfnamefont{M.}~\bibnamefont{Rabaud}}, \bibnamefont{and}
  \bibinfo{author}{\bibfnamefont{Y.}~\bibnamefont{Couder}},
  \bibinfo{journal}{Europhys.\ Lett.} \textbf{\bibinfo{volume}{22}},
  \bibinfo{pages}{17} (\bibinfo{year}{1993}).

\bibitem[{\citenamefont{Willaime et~al.}(1993)\citenamefont{Willaime, Cardoso,
  and Tabeling}}]{willaime1993}
\bibinfo{author}{\bibfnamefont{H.}~\bibnamefont{Willaime}},
  \bibinfo{author}{\bibfnamefont{O.}~\bibnamefont{Cardoso}}, \bibnamefont{and}
  \bibinfo{author}{\bibfnamefont{P.}~\bibnamefont{Tabeling}},
  \bibinfo{journal}{Phys.\ Rev.\ E} \textbf{\bibinfo{volume}{48}},
  \bibinfo{pages}{288} (\bibinfo{year}{1993}).

\bibitem[{\citenamefont{Degen et~al.}(1996)\citenamefont{Degen, Mutabazi, and
  Andereck}}]{degen1996}
\bibinfo{author}{\bibfnamefont{M.}~\bibnamefont{Degen}},
  \bibinfo{author}{\bibfnamefont{I.}~\bibnamefont{Mutabazi}}, \bibnamefont{and}
  \bibinfo{author}{\bibfnamefont{C.~D.} \bibnamefont{Andereck}},
  \bibinfo{journal}{Phys.\ Rev.\ E} \textbf{\bibinfo{volume}{53}},
  \bibinfo{pages}{3495} (\bibinfo{year}{1996}).

\bibitem[{\citenamefont{Colovas and Andereck}(1997)}]{colovas1997}
\bibinfo{author}{\bibfnamefont{P.~W.} \bibnamefont{Colovas}} \bibnamefont{and}
  \bibinfo{author}{\bibfnamefont{C.~D.} \bibnamefont{Andereck}},
  \bibinfo{journal}{Phys.\ Rev.\ E} \textbf{\bibinfo{volume}{55}},
  \bibinfo{pages}{2736} (\bibinfo{year}{1997}).

\bibitem[{\citenamefont{Bottin et~al.}(1978)\citenamefont{Bottin, Dauchot, and
  Daviaud}}]{bottin1997}
\bibinfo{author}{\bibfnamefont{S.}~\bibnamefont{Bottin}},
  \bibinfo{author}{\bibfnamefont{O.}~\bibnamefont{Dauchot}}, \bibnamefont{and}
  \bibinfo{author}{\bibfnamefont{F.}~\bibnamefont{Daviaud}},
  \bibinfo{journal}{Phys.\ Rev.\ Lett.} \textbf{\bibinfo{volume}{79}},
  \bibinfo{pages}{4377} (\bibinfo{year}{1978}).

\bibitem[{\citenamefont{Vallette et~al.}(1997)\citenamefont{Vallette, Jacobs,
  and Gollub}}]{gollub1997}
\bibinfo{author}{\bibfnamefont{D.~P.} \bibnamefont{Vallette}},
  \bibinfo{author}{\bibfnamefont{G.}~\bibnamefont{Jacobs}}, \bibnamefont{and}
  \bibinfo{author}{\bibfnamefont{J.~P.} \bibnamefont{Gollub}},
  \bibinfo{journal}{Phys.\ Rev.\ E} \textbf{\bibinfo{volume}{55}},
  \bibinfo{pages}{4277} (\bibinfo{year}{1997}).

\bibitem[{\citenamefont{Hinrichsen}(2000)}]{hinrichsen2000a}
\bibinfo{author}{\bibfnamefont{H.}~\bibnamefont{Hinrichsen}},
  \bibinfo{journal}{Adv.\ Phys.} \textbf{\bibinfo{volume}{49}},
  \bibinfo{pages}{815} (\bibinfo{year}{2000}).

\bibitem[{\citenamefont{Jensen}(1999)}]{jensen1999}
\bibinfo{author}{\bibfnamefont{I.}~\bibnamefont{Jensen}}, \bibinfo{journal}{J.\
  Phys.\ A} \textbf{\bibinfo{volume}{32}}, \bibinfo{pages}{5233}
  (\bibinfo{year}{1999}).

\bibitem[{\citenamefont{Grassberger}(1997)}]{grassberger1997}
\bibinfo{author}{\bibfnamefont{P.}~\bibnamefont{Grassberger}}, in
  \emph{\bibinfo{booktitle}{Nonlinearities in complex systems, proceedings of
  the 1995 Shimla conference on complex systems}}, edited by
  \bibinfo{editor}{\bibfnamefont{S.}~\bibnamefont{Puri}}
  (\bibinfo{publisher}{Narosa publishing}, \bibinfo{year}{1997}).

\bibitem[{\citenamefont{Rosensweig}(1985)}]{rosensweig}
\bibinfo{author}{\bibfnamefont{R.~E.} \bibnamefont{Rosensweig}},
  \emph{\bibinfo{title}{Ferrohydrodynamics}} (\bibinfo{publisher}{Cambridge
  University Press}, \bibinfo{address}{Cambridge}, \bibinfo{year}{1985}).

\bibitem[{\citenamefont{Mahr and Rehberg}(1998)}]{mahr1998}
\bibinfo{author}{\bibfnamefont{T.}~\bibnamefont{Mahr}} \bibnamefont{and}
  \bibinfo{author}{\bibfnamefont{I.}~\bibnamefont{Rehberg}},
  \bibinfo{journal}{Physica\ D} \textbf{\bibinfo{volume}{111}},
  \bibinfo{pages}{335} (\bibinfo{year}{1998}).

\bibitem[{\citenamefont{Friedrichs and Engel}(2000)}]{friedrichs2000}
\bibinfo{author}{\bibfnamefont{R.}~\bibnamefont{Friedrichs}} \bibnamefont{and}
  \bibinfo{author}{\bibfnamefont{A.}~\bibnamefont{Engel}},
  \bibinfo{journal}{Euro.\ Phys.\ J.\ B} \textbf{\bibinfo{volume}{18}},
  \bibinfo{pages}{329} (\bibinfo{year}{2000}).

\bibitem[{\citenamefont{Cross and Hohenberg}(1993)}]{cross1993}
\bibinfo{author}{\bibfnamefont{M.~C.} \bibnamefont{Cross}} \bibnamefont{and}
  \bibinfo{author}{\bibfnamefont{P.~C.} \bibnamefont{Hohenberg}},
  \bibinfo{journal}{Rev. Mod. Phys.} \textbf{\bibinfo{volume}{65}},
  \bibinfo{pages}{851} (\bibinfo{year}{1993}).

\bibitem[{\citenamefont{Kaneko and Tsuda}(2000)}]{kaneko}
\bibinfo{author}{\bibfnamefont{K.}~\bibnamefont{Kaneko}} \bibnamefont{and}
  \bibinfo{author}{\bibfnamefont{I.}~\bibnamefont{Tsuda}},
  \emph{\bibinfo{title}{Complex Systems: Chaos and Beyond}}
  (\bibinfo{publisher}{Springer and Berlin}, \bibinfo{year}{2000}).

\bibitem[{\citenamefont{Janssen et~al.}(1999)\citenamefont{Janssen, \"u.
  Kutbay, and Oerding}}]{janssen1999}
\bibinfo{author}{\bibfnamefont{H.~K.} \bibnamefont{Janssen}},
  \bibinfo{author}{\bibnamefont{\"u. Kutbay}}, \bibnamefont{and}
  \bibinfo{author}{\bibfnamefont{K.}~\bibnamefont{Oerding}},
  \bibinfo{journal}{J.\ Phys.\ A} \textbf{\bibinfo{volume}{32}},
  \bibinfo{pages}{1809} (\bibinfo{year}{1999}).

\bibitem[{\citenamefont{Bohr et~al.}(2001)\citenamefont{Bohr, van Hecke,
  Mikkelsen, and Ipsen}}]{bohr2001}
\bibinfo{author}{\bibfnamefont{T.}~\bibnamefont{Bohr}},
  \bibinfo{author}{\bibfnamefont{M.}~\bibnamefont{van Hecke}},
  \bibinfo{author}{\bibfnamefont{R.}~\bibnamefont{Mikkelsen}},
  \bibnamefont{and} \bibinfo{author}{\bibfnamefont{M.}~\bibnamefont{Ipsen}},
  \bibinfo{journal}{Phys.\ Rev.\ Lett.} \textbf{\bibinfo{volume}{86}},
  \bibinfo{pages}{5482} (\bibinfo{year}{2001}).

\bibitem[{\citenamefont{Manneville}(1990)}]{manneville}
\bibinfo{author}{\bibfnamefont{P.}~\bibnamefont{Manneville}},
  \emph{\bibinfo{title}{Dissipative Structures and Weak Turbulence}}
  (\bibinfo{publisher}{Academic Press and Boston}, \bibinfo{year}{1990}).

\bibitem[{\citenamefont{Kinzel}(1983)}]{kinzel1983}
\bibinfo{author}{\bibfnamefont{W.}~\bibnamefont{Kinzel}}, in
  \emph{\bibinfo{booktitle}{Percolation structures and processes}}, edited by
  \bibinfo{editor}{\bibfnamefont{G.}~\bibnamefont{Deutscher}},
  \bibinfo{editor}{\bibfnamefont{R.}~\bibnamefont{Zallen}}, \bibnamefont{and}
  \bibinfo{editor}{\bibfnamefont{J.}~\bibnamefont{Adler}}
  (\bibinfo{publisher}{Adam Hilger}, \bibinfo{year}{1983}),
  vol.~\bibinfo{volume}{5}, pp. \bibinfo{pages}{425 -- 445}.

\end{thebibliography}
\end{document}